\documentclass[letterpaper]{article} % DO NOT CHANGE THIS
\usepackage[]{aaai24}  % DO NOT CHANGE THIS
\usepackage{tabularx}
\usepackage{times}  % DO NOT CHANGE THIS
\usepackage{helvet}  % DO NOT CHANGE THIS
\usepackage{courier}  % DO NOT CHANGE THIS
\usepackage[hyphens]{url}  % DO NOT CHANGE THIS
\usepackage{graphicx} % DO NOT CHANGE THIS
\usepackage{natbib}  % DO NOT CHANGE THIS AND DO NOT ADD ANY OPTIONS TO IT
\usepackage{caption} % DO NOT CHANGE THIS AND DO NOT ADD ANY OPTIONS TO IT
\usepackage{amssymb}
\usepackage{algorithm}
\usepackage{algorithmic}

\usepackage{newfloat}
\usepackage{listings}
\DeclareCaptionStyle{ruled}{labelfont=normalfont,labelsep=colon,strut=off} % DO NOT CHANGE THIS
\floatstyle{ruled}
\newfloat{listing}{tb}{lst}{}
\floatname{listing}{Listing}
\title{Generative AI Needs Adaptive Governance}
\author{
    Anka Reuel\textsuperscript{\rm 1}\equalcontrib,
    Trond Arne Undheim\textsuperscript{\rm 1}\equalcontrib
}
\affiliations{
    \textsuperscript{\rm 1}Stanford University\\
    anka@cs.stanford.edu, trondun@stanford.edu
}

\usepackage{bibentry}
\begin{document}

\maketitle

\begin{abstract}
Because of the speed of its development, broad scope of application, and its ability to augment human performance, generative AI challenges the very notions of governance, trust, and human agency. The technology’s capacity to mimic human knowledge work, feedback loops including significant uptick in users, research-, investor-, policy-, and media attention, data and compute resources, all lead to rapidly increasing capabilities. For those reasons, adaptive governance, where AI governance and AI co-evolve, is essential for governing generative AI. In sharp contrast to traditional governance’s regulatory regimes that are based on a mix of rigid one-and-done provisions for disclosure, registration and risk management, which in the case of AI carry the potential for regulatory misalignment, this paper argues that generative AI calls for adaptive governance. We define adaptive governance in the context of AI and outline an adaptive AI governance framework. We outline actors, roles, as well as both shared and actors-specific policy activities. We further provide examples of how the framework could be operationalized in practice. We then explain that the adaptive AI governance stance is not without its risks and limitations, such as insufficient oversight, insufficient depth, regulatory uncertainty, and regulatory capture, and provide potential approaches to fix these shortcomings.
\end{abstract}

\section{Introduction}

%What is the problem?
%Why is it interesting and important?
%Why is it hard? (E.g., why do naive approaches fail?)
%Why hasn't it been solved before? (Or, what's wrong with previous proposed solutions? How does mine differ?)
%What are the key components of my approach and results? Also include any specific limitations.

Generative AI, like GPT-4, is already capable of producing convincing real-world content, including text, code, images, music, and video, based on vast amounts of training data \cite{Feuerriegel2023}. It automates and scales up information processing, significantly augmenting human creative and cognitive expression. As a result, generative AI not only tangibly impacts the tasks (knowledge-)workers perform, but also impacts the workflow and the quality of work, transforming the human-AI relationship \cite{Pflanzer2023}. The impacts seem to extend even to advanced office work conducted by highly educated, skilled knowledge workers with years of experience \cite{DellAcqua2023}. \\

For these reasons, generative AI is rapidly evolving into a general-purpose technology that may have a distinct, aggregate impact across  industry and society, possibly in short order. Because of the speed of improvement, an emergent AI anxiety vexed even AI experts \cite{Heaven2023a}, and the awareness for the capabilities and risks of these AI systems increased across technical experts, policymakers, and the general public. However, the inherent complexity and speed of what is going on challenges AI policymaking. Major governmental actors such as China, the EU, the UK, and the US have taken notice, and are formulating or updating their AI policy initiatives to try to effectively govern the new technology \cite{Roberts2023, WH2023}. One-and-done regulatory regimes based on a mix of disclosure, registration, and risk management provisions are being discussed \cite{euaiact2024, sheehan2023china, us2024executiveorder}, each carrying the potential for regulatory misalignment, where proposals may distract, fail, or backfire \cite{Lawrence2023}. As a whole, those proposals and responses can be characterized as fast-fix sound bites caught in slow-moving legislative processes and top-down implementation. In short, they are traditional governance responses.\\

%expert-driven policy development,
In this paper, we argue that such a traditional approach is not sufficient to deal with generative AI. We specifically explore what is different about generative AI that warrants an adaptive governance approach and how such an approach could look like. We make the following contributions:
\begin{itemize}
    \item We define traditional and adaptive approaches to AI governance
    \item We justify the need for adaptive AI governance for generative AI
    \item We define a framework for adaptive AI governance and show examples of how it can be operationalized
    \item We outline the limitations of such an approach and provide suggestions what can be done to overcome them
\end{itemize}

The paper is structured as follows: In Section 2, we explore in more detail what's different about generative AI that warrants an adaptive governance approach. We then describe in Section 3 the traditional AI governance approach, before we explain the adaptive AI governance approach in Section 4. We describe a general framework for  such an approach in Section 5 and provide examples how it can be operationalized in Section 6. We conclude with Section 7, where we outline downsides and limitations of such an adaptive AI governance approach.

\section{What's Different About Generative AI}

In 2017, researchers found that generative pre-trained transformers (GPTs) could simplify language model tasks such as machine translation and text summarization \cite{Wang2023, vaswani2017attention}. Since then, progress has been rapid \cite{aiindex2024}. In 2022, OpenAI’s chatbot, ChatGPT, brought generative AI to the general public. OpenAI managed to (1) scale GPTs, making them capable of addressing a wide range of requests by users, which made them seem to have some level of intelligence \cite{AlLily2023} and (2) provide non-technical people with an interface to interact with AI and experience it first-hand. With that, AI moved from an abstract concept and expert tool, to something that everyone could play around with and use.\\

As a results, ChatGPT and similar AI systems like Claude by Anthropic \cite{anthropic2023claude} have seen unprecedented adoption \cite{Hu2023} as a supplement to search engines, turning searches into conversations \cite{Stokel-Walker2023}, transforming customer conversations \cite{Ramesh2022}, boosting digital content production \cite{Davenport2022}, and potentially laying the foundations for a metaverse \cite{Lv2023}. Other, more inconspicuous use cases, such as patient triage, risk management, songwriting, test preparation, and voice acting are emerging \cite{Ray2023}, in part due to the systems' increasing multi-modal capabilities \cite{aiindex2024}.\\ 

Immediate risks of such applications include copyright infringement \cite{Appel2023, Brittain2023, Murray2023}, exposure of (sensitive) training data \cite{Carlini2020}, and impersonation \cite{tariq2022real}, while longer term concerns include uncontrolled (‘unaligned’) artificial general intelligence \cite{Turchin2019, Jungherr2023}. However, given the general-purpose nature of generative AI – and foundation models in particular –, and their increased usage across a wide variety of applications and contexts, new risks are constantly emerging, manifesting in an increase in harms and incidents \cite{aiindex2024}.\\

Within the ranks of big tech, OpenAI’s initial success with ChatGPT created a race to the top and has meant the tech industry has put far greater resources behind generative AI than they did for previous generations of AI \cite{GoldmanSachs2023}. Sector- and firm-specific versions of generative AI models outside big tech are also in rapid development. Notable examples include finance, healthcare, life sciences, and marketing with their respective associated risks \cite{bommasani2021opportunities}. Big companies that have made major generative AI announcements include AWS, Google, IBM, Microsoft, NVIDIA, OpenAI, and Oracle \cite{Leone2023}, and notable startups working in the fiels include OpenAI, Anthropic, Synthesis AI, and Stability AI \cite{Hiter2023}. In addition, a rise of perceived market value and opportunities of generative AI led to significantly increased investments and talent attraction. AI investment is forecast to approach \$200 billion globally by 2025 and the AI market is expected to grow twentyfold by 2030, up to nearly two trillion U.S. dollars \cite{GoldmanSachs2023, Roser2023, Thormundsson2023}.\\ 

Simultaneously, AI research and development (R\&D) has also skyrocketed. There has been a significant increase in AI papers at computer science conferences, many of them from industry due to the associated costs and required compute, given the size of the models \cite{aiindex2024}. The research explores both improvements and extensions to current models, as well as risks and potential mitigation strategies. These shifts make it hard to keep up with research trends because they happen extremely quickly, even for academics in the field \cite{Arnold2023b}. In addition, given the speed of development of generative AI, scientific uncertainty is exacerbated \cite{Wallach2019}. Generative AI can itself be used to make scientific discoveries, such as in the case of AlphaFold’s protein predictions, and has yet uncertain drug discovery potential \cite{Arnold2023a}.\\

%Besides an increase in attention to these models by industry, we've also seen an increase in public attention, fueled by better accessibility and 

%, both in academic and non-academic positions, which in turn fuels research and innovation in the AI space

Increased attention has shaped the public and media's understanding and attitudes towards AI. The risks of particularly generative AI are now a topic high on the agenda. There are calls for policymakers to prioritize AI governance with urgency, although there is disagreement about whether the most significant risks that need to be addresses are current \cite{Nature2023}, emergent \cite{OBrien2020}, or future risks \cite{Price2023}, and whether to pursue seemingly quick fixes such as centralized oversight bodies, which China has done \cite{Cheng2023}, pause AI development \cite{Metz2023} or shut it down \cite{Yudkowsky2023} until we figure it all out. This has led to the worry that policies don’t show the needed sophistication, that bad policies would backfire, and that risk mitigation approaches might become counterproductive \cite{Hagendorff2023}.\\

The broad economic implications of generative AI are also uncertain. The estimated annual value of generative AI to the US GDP is \$1.043 trillion by 2032 \cite{cognizant2024}. Predictions for job market impacts include wide-spread job alterations and worker displacements; for example, one study predicts that as much as 90\% of jobs will be impacted by generative AI, and 9\% of worker will be displaced by the technology \cite{cognizant2024}.\\

To summarize, generative AI is different or surpasses previous generations of AI in the following ways:
\begin{itemize}
    \item It's more accessible to technical and non-technical users, significantly increasing its adoption across a wider user population
    \item It's at least as complex as previous AI, given the increasing size and complexity of the models
    \item It's significantly more expensive, pricing independent actors like academia out of research and independent scrutiny
    \item The speed of research in the field is at unprecedented levels
    \item It's predicted to have an exceeding impact on the workforce and the economy
\end{itemize}

These characteristics raise the following issues that are key for policymakers to understand in the context of generative AI governance: (1) understanding the capabilities and limitations of generative AI requires extensive technical background knowledge, (2) this expertise and state-of-the-art knowledge mostly resides with industry players that have their own interests, the reason being that massive computing resources are necessary often inaccessible to academia \cite{aiindex2024}, and (3) the rapid speed of development and research often causes expert knowledge to be outdated or irrelevant within months.

\section{Traditional AI Governance}

Legislative stability is a staple of political thought and seen as preconditions of due process, the quality of law making, and legal certainty \cite{Sebok2022}. As a result, today, policymakers often treat laws as a one-off thing where you write a law and then don’t touch it for another five to ten years (or ever). This causes a significant mismatch between the speed of new developments and discoveries in AI versus the speed of legislative processes, creating a ``pacing-problem'' of ``cumbersome procedural and bureaucratic procedures and safeguards'' \cite{Wallach2019} where regulatory systems ``fail to put in place appropriately tailored regulatory measures by the time new applications of fast-moving technologies begin to affect society'' \cite{Wallach2019}.\\

Traditional approaches to governance are characterised by ``top-down directives or
command-and-control policies'' \cite{chaffin2014decade} and inflexibility with respect to changing, uncertain situations, along with a tendency to ``fall
short in efforts to coordinate governance across large-scale ecosystems that cross multiple jurisdictional boundaries'' \cite{lemos2006environmental, chaffin2014decade}.\\

Take the EU AI Act, for example. It was first proposed by the European Commission on 21 April 2021 and approved in March 2024, and it will come into effect between six and 36 months after it has officially been published \cite{euaiact2024, Consilium2023}. While some adjustments have been made in the past three years, it will likely be an emergent case study in attempting to regulate a rapidly changing emergent technology using traditional regulatory mechanisms. The AI world was vastly different three years ago; transformers were just discovered in 2017, AlphaFold was yet to be released, ChatGPT didn't exist yet, and generative AI was not a widely known term. During the process of designing the EU AI Act, foundation models took off and had an increasing impact across sectors. However, foundation models did not fit – and hence could not be addressed – in the EU’s original risk-tier approach to AI because they only enable downstream applications but in themselves don’t necessarily cause direct harm (yet); hence, mid-way through the legislative process, an extra clause had to be added to account for these models. Such a change was only possible because the negotiations for the law were ongoing; if foundation models would have taken off post-enactment of the AI Act, they may have rendered it largely ineffective for this type of AI.\\

Another example of traditional governance in the context of the EU AI Act is the classification of systems with systemic risk. The EU will measure systemic risk of general purpose AI (GPAI) models, outside of policy circles more commonly known as foundation models, and other AI systems based on how much computation it took to train these systems. The metric used is whether the training required greater than $10^25$ floating-point operations per second (FLOPS), where FLOPS is a measure for how many arithmetic operations a computer can perform in one second. While the exact number hasn’t been disclosed, it is estimated that GPT-4 handled on the order of $10^25$ FLOPS of operations. Using that threshold makes sense today, but what if improvements in algorithmic approaches drastically reduce the need for FLOPS but retain advanced capabilities?  For example, the regulation could have added language on adjustments to the metric over time, or specified criteria that would have triggered revisiting the definition of systemic risk. Moreover, reliance on a single metric gives some level of regulatory certainty for vendors but is not sufficient to ensure safety because it encourages experimentation around the $10^25$ threshold and will simply split models into several distinct products trained to near threshold. Yet, the criteria is part of the final regulatory text, without provisions to update or revisit it.\\

%Adding to the complexity, nation states, and economic area organizations such as the EU cannot truly be seen as impartial on AI because of the economic stakes of leadership of  general-purpose technology development and the industrial applications to follow. Additionally, corporate AI governance is uneven, immature, non-consolidated, and deploys non-interoperable frameworks \cite{Cihon2021, Mantymaki2022, Mokander2022}.\\ 

%There further is a tendency to rely on potentially biased information from the media and industry. As a result, it is hard for policymakers to build their own opinion on the capabilities and limitations of generative AI or what policies might be sensible.

\section{Adaptive AI Governance}

In contrast, adaptive governance is a concept that originated in the environmental governance space ``for the holistic management of complex environmental problems'' \cite{sharma2018adaptive}(see also \citet{dietz2003struggle}, \citet{folke2005adaptive} and \citet{walker2004resilience}), following ``the failure of previous management regimes to implement governance structures robust enough to achieve ecological sustainability and build community capacity under conditions of uncertainty'' \cite{sharma2018adaptive}. It's characterized by ``flexibility, resilience, and capacity for change in the planning and implementation process'' \cite{sharma2018adaptive}.\\

%Adaptive governance has emerged in the last decade as an intriguing avenue of theory and practice for the holistic management of complex environmental problems (Dietz et al., 2003; Folke et al., 2005; Walker et al., 2004). In response to the failure of previous management regimes to implement governance structures robust enough to achieve ecological sustainability and build community capacity under conditions of uncertainty, adaptive governance expressly refuses a narrow focus on linear management for ecosystem outcomes (Brunner et al., 2005; Pahl-Wostl, 2009). The field instead reflects a growing need for environmental and resource management regimes which include concepts of flexibility, resilience, and capacity for change in the planning and implementation process. https://www.sciencedirect.com/science/article/pii/S030147971830598X

%Adaptive governance must therefore contemplate a level of flexibility and evolution in governmental action beyond that currently found in the heavily administrative governments of many democracies. https://www.ncbi.nlm.nih.gov/pmc/articles/PMC5954422/

Adaptive governance is – ideally – fast, flexible, responsive, and iterative \cite{Janssen2020}. Considering the uncertainty, it must simultaneously still be deliberative \cite{Nordstrom2022, Innerarity2023}. For that reason, it is informed by normative policy shapers \cite{Smith2023}, and learning is a key value \cite{Janssen2016}, alongside firm but gentle coordination of stakeholders and policies that results in comprehensive monitoring \cite{Wallach2019}. At best, adaptive governance is an approach where learning from change makes the governance model better and better so it never risks being outdated. Adaptive governance has been successfully implemented in other domains, such as public health around COVID-19 \cite{Janssen2020, Khan2021}, and sustainability governance in the context of climate change \cite{Schultz2015, Linkov2018, May2022, Mourby2022}.\\

There are also parallels with the condition of governing under uncertainty in other evolving technologies. One could think of nuclear technology governance as it evolved in the post World War II era \cite{Wu2019, Khlaaf2023}, biotech regulation starting in the 1980s \cite{Huzair2021, Trump2022}, and nanotech regulation in the 2000s \cite{Wolinsky2006, Guston2014, Allan2021}. Each emerging technology has different characteristics: nuclear was stable until the recent small, modular reactors \cite{Sam2023} and its second phase of proliferation in emerging markets \cite{Wu2019}. Biotech (and its regulation) has been constantly changing for 40 years \cite{Li2021, Sheahan2021, Mourby2022}, and significant nanotech breakthroughs coupled with obscure and scattered risk governance tools continuously challenge nano regulation efforts \cite{Mullins2022}. Yet, they share parallels that inform our idea of adaptive governance under uncertainty.\\

Adaptive governance initiatives are typically coupled with ways to share good approaches with others in the regulatory network. This has been successfully done with e-government implementation case studies across the EU \cite{EU2023}. However, just having a best practice framework for AI governance is not enough; being able to share, learn, and get inspired from each other might avoid crucial mistakes, and might stop agencies from reinventing the wheel and not wasting valuable resources.\\

Adaptive governance in the context of policy making in the digital realm takes inspiration from the principles of agile methodology \cite{} which originated in software development, and emphasizes adaptability, stakeholder collaboration, and rapid response to change. The idea is that software policy needs to be flexible, responsive, and iterative because that’s how software is created, implemented, and modified. Specifically for adaptive AI governance, the approach also needs to be evolutionary and social in nature, plus incorporate solid processes for AI-human collaboration \cite{Caldwell2022}. Governance must also match and mimic the iterative development, speed, and collaboration patterns of the generative AI development process \cite{Feuerriegel2023}. Adaptive AI governance \cite{Agbese2023}, just like previous generations of adaptive IT governance, must be ambidextrous \cite{Janssen2016}; in other words, policy organizations should be able to pursue potentially contradictory aims at the same time. For example, it should be possible to both emphasize close monitoring and yet allow for significant, disruptive innovation through careful anticipatory governance \cite{Guston2014, Heo2021}, allowing for uncertainty. If and when operating under particularly high uncertainty, there is the need to engage in, and feel confident about, the balancing act of tentative governance \cite{Kuhlmann2019}, testing things out and revising quickly if measures prove counterproductive. The United Nations Development Program (UNDP) describes such an approach as triple-A governance, both anticipatory, agile, and adaptive \cite{Ramos2020}.\\ 

Adaptive (AI) governance takes larger systems with impact on other industries and on society into account. There is early evidence that generative AI might upend even food, agriculture, financial services, mining, and telco industries across the manufacturing supply chain \cite{Freire2023, Xu2022, Ebni2023, Jan2023}, increasing the need for said adaptive governance. A variety of stakeholders are included, not just one organization and its internal response. Subnational dynamics also cannot be ignored \cite{Liebig2022}.\\

To treat laws and regulations as a one-off thing only works in relatively mature industries, where developments are comparatively slow, predictable, and changes don’t happen as frequently. As we've shown in Section 2, none of these characteristics apply to the field of (generative) AI yet. In addition, generative AI has two layers to it: a context-dependent one where impacts of the technology manifest in specific application contexts only, and a non-context-dependent layer that has an impact across sectors, complicating the governance process as the number of application contexts constantly increases while the impact of the non-context-dependent layer remains uncertain. Hence, adaptive governance is better suited for generative AI: It is able to keep pace with the technology's rapid development because it is faster, which reduces the pacing problem, it is more flexible which allows to easier address changes in technological capacities, and it has built-in feedback loops, which enables policymakers to proactively react to (in-)efficiencies, loopholes, or misguided legal provisions. The latter is specifically relevant since it remains an open question which governance measures are effective in the context of (generative) AI.

%Governance approaches need to address both to be effective. Generative AI transcends traditional sector boundaries, necessitating a holistic approach. 

Summarizing the differences between them, traditional AI governance is slow, static and only encompasses a small set of actors. In contrast, adaptive AI governance is fast, dynamic, and encompasses a large set of actors (see Table \ref{tab:traditionalvsadaptive}).

\begin{table}[]
    \centering
    \begin{tabular}{|l|l|l|}
    \hline
    & \textbf{Traditional} & \textbf{Adaptive} \\
    \hline
    \textbf{Speed} & Slow & Fast \\
    \hline
    \textbf{Content} & Static & Dynamic \\
    \hline
    \textbf{Actors} & Small set & Large set \\
    \hline
    \end{tabular}
    \caption{Traditional vs. Adaptive AI Governance}
    \label{tab:traditionalvsadaptive}
\end{table}

\section{A Framework for Adaptive Governance in AI}

What are potential frameworks or approaches that foster adaptive governance in the field of AI? Soft laws might be one suggested approach, as they can help prepare the road to binding laws and act as test beds for what may work \cite{reuel2024softlaws}. They might signal best practices that we can incorporate into binding regulations. However, they are not sufficient. We need to get to an adaptive governance structure that ensures a high degree of bindingness while remaining flexible to future developments and changes. To put it differently, adaptive governance would mandate the implementation of reasonable (minimum) safety and responsibility standards, given the current state of knowledge, but remain flexible to iteratively improve based on new insights.\\

In this chapter, we're adopting the adaptive governance approach from environmental governance to the field of AI, and expand it based on insights from other fields. Building and adopting such a framework means specifying relevant actors, tasks, roles, and activities; while specific instantiations of the adaptive AI governance framework will be context-dependent,   we provide specific examples of how adaptive AI governance could look in practice below. 

%An adaptive AI governance framework would entail a mix of \textit{Monitoring} (system and people), \textit{Training} (employees and ecosystem), \textit{Governance R&D} (AI for governance, foresight), and \textit{Safety Management} encompassing \textit{Environment, Health and Safety} (EHS).\\ 
%Any good framework must help us act and cannot just inform us \cite{Fazey2020}.
We'll start with defining the set of actors that would fall within the scope of an adaptive AI governance approach: In a traditional triple Helix innovation model, which describes interactions between actors in a society \cite{Cai2022, Leydesdorff1997, Carayannis2012}, we have government, industry and academia as the three essential stakeholders. Combining Helix innovation theory with insights from actor network theory \cite{Edwards2022, Morton2023}, we derive an expanded innovation model for (generative) AI that encompasses an extended set of actors to be considered in a governance framework: governments, industry, academia, civil society, and citizens. Adaptive governance is then about the co-governance of these different actors. In this adaptive governance framework, by default, all actors are equally important or at least to the extent that no actor can be reduced to another for any purpose \cite{Fornazin2016}.\\
%environment
%Human-AI hybrid actions are the collaborative interworking of human agents and AI-enabled systems \cite{Fabri2023}. For example, any decision created through the use of an AI system could be categorized as a hybrid action, but there are important degrees of sociomateriality, with more or less human or material (machine/AI) input into the deliberations and actions that ensue \cite{Fabri2023}. However, without ultimately being represented by humans, AI systems cannot fit within the current legal responsibility model.\\

%In a basic sense, being designed by humans, AI already reflects human activity \cite{Taylor2023}. That being said, governing by aligning AI to human values \cite{Choung2023} is too limited, and may not be possible in the first place, given the plurality of values held among humans. A related challenge is the lack of a consolidated set of AI ethics frameworks \cite{Qiang2023}. Instead, understanding systemic properties across and between stakeholders is equally important for effective governance. 

Secondly, we define the high-level tasks that need to be fulfilled in the context of an adaptive AI governance framework. We split them into general (\textit{SCUMIA}, the acronym of the shared activities) and actor-specific activities (the \textit{FACTI} activities), as indicated in Table \ref{tab:tasks}.

\begin{table*}[]
    \centering
    \begin{tabularx}{0.80\textwidth}{|X|c|c|c|c|c|X|}
        \hline
        & \textbf{Government} & \textbf{Industry} & \textbf{Academia} & \textbf{Civil Society} & \textbf{Citizens}\\
        \hline
        \textbf{Share Best Practice} & \checkmark & \checkmark & \checkmark & \checkmark & \checkmark\\
        \hline
        \textbf{Collaborate} & \checkmark & \checkmark & \checkmark & \checkmark & \checkmark \\
        \hline
        \textbf{Use} & \checkmark & \checkmark & \checkmark & \checkmark & \checkmark  \\
        \hline
        \textbf{Monitor} & \checkmark & \checkmark & \checkmark & \checkmark & \checkmark \\
        \hline
        \textbf{Inform} & \checkmark & \checkmark & \checkmark & \checkmark & \checkmark \\
        \hline
        \textbf{Adapt} & \checkmark & \checkmark & \checkmark & \checkmark &  \checkmark \\
        \hline
        \hline
        \textbf{Finance} & \checkmark & \checkmark &  & \checkmark &  \\
        \hline
        \textbf{Anticipate} & \checkmark & \checkmark & \checkmark &  &  \\
        \hline
        \textbf{Challenge} &  &  & \checkmark & \checkmark &  \\
        \hline
        \textbf{Train} & \checkmark & \checkmark & \checkmark & \checkmark &  \\
        \hline
        \textbf{Innovate} &  & \checkmark & \checkmark &  &  \\
        \hline
    \end{tabularx}
    \caption{Activities of different actors in an adaptive governance framework}
    \label{tab:tasks}
\end{table*}

The shared \textit{SCUMIA} activities include
\begin{itemize}
    \item \textbf{Share Best Practices}: Actors should openly share lessons learned, successful approaches, and innovative ideas to foster collective learning and improvement in AI governance. This could involve publishing case studies, participating in multi-stakeholder forums, etc.
    \item \textbf{Collaborate:} All actors should actively work together on initiatives to advance responsible AI development and deployment. This may include joint research projects, public-private partnerships, and other collaborative efforts that leverage the strengths of each group.
    \item \textbf{Use:} All actors should strive to adopt and implement best practices and insights from the adaptive governance process. Academia and civil society should utilize available tools to study and provide insights on AI systems. Industry and government should put key learnings into practice.
    \item \textbf{Monitor:} There needs to be ongoing monitoring by all parties to assess the effectiveness of current AI governance measures and identify areas for improvement. This could involve tracking key metrics, conducting impact assessments, and gathering feedback from affected stakeholders.
    \item \textbf{Inform:} Sharing of information is critical to enable evidence-based decision making. All actors should proactively communicate developments, concerns, and opportunities pertaining to (generative) AI and AI governance through appropriate channels.
    \item \textbf{Adapting:} All actors must be willing to adapt their approaches based on new learnings and changing circumstances. Governance structures need to be living documents, with clear processes for iteration and amendment as the (generative) AI landscape evolves.
\end{itemize}

In addition to the shared \textit{SCUMIA} activities, some actors would also be involved in actor-specific activities: \textit{Financing}, \textit{Anticipating}, \textit{Challenging}, \textit{Training}, and \textit{Innovating} (the \textit{FACTI} activities).

\begin{itemize}
    \item \textbf{Financing:} Government, industry and civil society should provide funding for AI governance initiatives, such as research on AI safety and ethics, public education campaigns, and multi-stakeholder collaborations. Innovative funding models like impact investing could be explored.
    \item \textbf{Anticipating:} Government, industry and civil society should engage in strategic foresight to anticipate future challenges and opportunities related to AI. This could involve trend analysis, scenario planning, and risk assessment to inform proactive policy and strategy.
    \item \textbf{Challenging:} Academia, civil society, and citizens play a vital role in constructively challenging the AI governance system to address shortcomings and spur continuous improvement. This could involve advocacy, critical analysis, and serving as watchdogs.
    \item \textbf{Training:} Actors like government, industry, academia and civil society should actively build AI governance capacity through education and skills development initiatives. This could target policymakers, developers, students, and the general public. 
    \item \textbf{Innovating:} Industry, and academia are uniquely positioned to drive AI (governance) innovation by developing new  technical tools or risk mitigation approaches to support AI governance \cite{reuel2024importance}.
\end{itemize}

Financing will largely be carried out by the government, industry, and civil society. Anticipating change is a shared task for government, industry, and civil society. Challenging the system is a shared responsibility between academia and civil society. Training others falls on government, industry, academia and civil society, while innovating largely involves industry and academia.\\

We explain how these activities would map to the different actors below, expanding on Table \ref{tab:tasks}.

\subsection{The role of government}

\begin{itemize}
    \item Governments have an expanded role in terms of anticipating change and adapting to it, as they need to future-proof and adapt (binding) AI governance measures to new and anticipated developments in the field.
    \item Governments have a duty to share best practices across society to/from all stakeholders and should focus on skills development and building a shared, digital commons (rules, heuristics, models, best practices, routines, ethical principles, data, interoperability principles/practices) in order to build a common pool of resources (‘pasture’) where benefits and harms are fairly distributed across all users \cite{Kreienkamp2020}. 
    \item The government’s role in AI governance is as facilitator and convenor, steering actors towards the public good and long term resilience to benefit the whole system, not as the ultimate judge.
    \item Governments need to further hire AI talent and build internal AI capacity, but also need to disclose any knowledge they have of AI being used within their jurisdiction and in markets regulated by them. 
    \item Governments should be accountable for AI safety at the national and regional levels, and should produce regular reports. 
\end{itemize}

\subsection{The role of industry}

\begin{itemize}
    \item Industry should focus on their corporate AI governance efforts to rapidly mature and consolidate them and to deploy interoperable frameworks that can be jointly monitored, tested, and updated as AI systems evolve. 
    \item Corporate boards of companies using AI at the core of their business should be mandated (or self-regulate) to have at least one board representative who is explicitly representing the responsible AI perspective, and that person should be armed with reporting tools to monitor progress and potential safety issues. 
    \item AI industry leaders need to implement strong cross-industry AI governance initiatives that are standardized, interoperable, transparent, and committed to safety. 
    \item The Chief Risk Officer (CRO) role needs to evolve into a more senior role and requires proper governance tools shared across the industry, also to be embedded in a national network of CROs.
    \item Cross-industry risk surveys cannot be voluntary to respond to but must be made mandatory.
\end{itemize}

\subsection{The role of academia}

\begin{itemize}
    \item Academia should dedicate time and resources explaining AI to policy makers, e.g., through training sessions, in op-eds, media interviews, and in crossover journals. 
    \item Academia needs to partner with other actors (e.g., industry or governments) to access sufficient computational resources to contribute to state-of-the-art (generative) AI research, development, innovation, and scrutiny.
\end{itemize}

\subsection{The role of civil society}
\begin{itemize}
    \item Civil society needs to build credible third-party expertise that can continuously question the status quo. Very few existing actors can do this alone, so networks are crucial. 
    \item Civil society should help bring attention to the societal implications of AI and advocate for responsible development and deployment practices by running public awareness campaigns, engage with media, and mobilize grassroots support for AI governance issues.
    \item Civil society should work to ensure that the perspectives of marginalized and vulnerable communities are considered in AI governance processes. They can conduct inclusive stakeholder consultations and advocate for policies that protect human rights and promote social justice. 
\end{itemize}

\subsection{The role of citizens}
\begin{itemize}
    \item Citizens should actively participate in debates surrounding the role of AI in government, in the workplace, at home, and in educational institutions. To do so, they need to stay informed and educate themselves, along with demanding intuitive AI interfaces not requiring special knowledge.
    \item Citizens should further inform other actors, such as civil societies or governments, about specific harms of AI systems they encountered.
\end{itemize}

\section{Concrete Examples of Adaptive AI Governance}

The next step in implementing an adaptive AI governance approach based on the framework above would be to clarify how these roles and activities can be operationalized. While a comprehensive, implementable adaptive governance framework is context-dependent because it needs to specify necessary budgets, parameters, data, indicators, emphasis, and processes that are shaped on a country's specific circumstances and priorities, we provide examples for adaptive AI governance in practice in this chapter. That being said, not all AI governance measures need to be made adaptive all at once; a first step could be to start with pilot programs where a select few measures are being implemented adaptively, with in-built feedback and adaptation loops into policies, and then governments could learn from these processes.\\

One previously suggested idea of governance coordinating committees (GCCs) situated outside government but with participation from government representatives, industry, nongovernmental organizations, think tanks, and other stakeholders \cite{Wallach2019} is one idea for operationalizing adaptive AI governance. However, we propose that such committees should instead be placed inside the government; each committee should have permanent government representatives that focus on AI and then invite external experts from a variety of stakeholders for regular debate and review of technological progress and potential adaptation of regulations. These deliberations should be given significant weight in the regulatory process. To address the dual-layer nature of the impact of AI, we should have both, dedicated government representatives at sector-specific government agencies while also a central AI officer (or similar) that is tasked with cross-sectoral impacts and developments of AI.\\

Another example of practical adaptive AI governance would be initial AI regulation that would include a passage that allows for a shorter legitimization process of new requirements based on the committees’ recommendations (suggested above). Another option in the context of AI regulations could be to build in structured, pre-determined revision and update rounds of the regulation that the suggested committees would lead.\\

Beyond that, there is a need to invest in regulatory AI governance R\&D, because the subject matter is complex, e.g., through building a national AI research resource, akin to advances by the US \cite{nairr2024us} and the UK \cite{airr2024uk}. In fact, such R\&D needs as significant attention as AI development itself. A percentage (for example three percent) of AI investment mandated towards AI governance and safety, matching the national R\&D to GDP ratio, which in the U.S. is 3.40 percent \cite{NCSES2021}, could be a reasonable target in an adaptive AI governance framework.\\

Additionally one could create centralized repositories and mandate organizations to register AI incidents and AI development, to be able to have an objective database that can be used to oversee current practices and extrapolate trends from current trajectories. These resources could then be used to assess and attempt to anticipate how AI might reshape key regulated and unregulated sectors (and sector governance) in the years ahead and build governance to that spec, not to the current state. To expand on this, governments should additionally build and debate potential future scenarios and corresponding governance responses in multi-stakeholder sessions.\\

In implementing an adaptive AI governance framework, one cannot ignore the short term risks of AI, such as bias, equity, and justice \cite{Agbese2023, Hagendorff2023, Nature2023, Ulnicane2023} in the desire to protect humanity from artificial general intelligence (AGI), an extremely ambiguous term for general purpose, high achieving cognitive technologies \cite{Heaven2023b}. Instead, we need to establish structures that allow for an objective view on risks from AI, both current and future, and design governance measures that address both current risks while anticipating future ones. Besides the above-mentioned national AI research resources and repositories/registers, governments could make (financial/compute) resources available to independent experts, e.g., academia or convened experts groups, akin to the IPCC but on a national scale and for AI, that could research and assess risks from AI systems independent of vetted interests (which is a concern with industry, along with associated regulatory capture, see \citet{dal2006regulatory}).\\

%Trust/mistrust goes and must go both ways, also between machines and humans \cite{Caldwell2022, Zanotti2023}. Since human experts tend to be overconfident \cite{Fabri2023}, and machines are only as careful as the AI models they use tells them to be, “trust and verify” procedures must also go both ways. However, AI governance should not aim for closure by command-and-control but rather choose a non-deterministic approach, using a flexible framework to nudge a process in a particular direction \cite{Kuhlmann2019} or unearth certain insights that might direct future, more certain, directional, or deterministic guidance.\\

Finally, if AI systems become intertwined with regular day-to-day activities of citizens, such as finance or healthcare, they cannot require specialty expertise to operate, monitor, or understand. Adaptive AI governance should focus on increasing the accessibility for and AI literacy of citizens by counteracting deliberate black boxing of AI \cite{Rudin2019}, a challenge with many facets \cite{Brozek2023}, and on implementing AI education initiatives, e.g., as part national school curricula or by providing educational resources for adults to help them understand the capabilities and limitations of generative AI. The goal of such an adaptive AI governance measure would be to empower citizens to use AI responsibly and to partake in adaptive governance efforts, especially if they are negatively impacted by the technology.

%The closest analogy we currently have for advanced systems with clever user interfaces are the no-code systems (anonymized, 2022) XXX available in enterprise software today. But evolving user interfaces to that level of simplicity gets infinitely more difficult the more advanced the underlying technologies are.

\section{Downsides of Adaptive AI Governance}
%Make not on limitations, esp. international
%\subsection{The role of human-AI hybrids}
%\begin{itemize}
    %\item In theory, hybrids need to be accountable for their actions the same way as other humans or workers are. In fact, they should have even higher bars for justifying their processes leading up to important actions with systemic implications. Defining relevant sub-categories of hybrids and hybrid actions will be key for both design of AI and for enforcement of AI governance \cite{Borsci2023}.
%\end{itemize}
Adopting an adaptive AI governance stance is not without its risks and limitations, notably insufficient oversight, insufficient depth, regulatory uncertainty, and regulatory capture.\\

Adaptive governance, with its rapid iteration and flexibility, may lead to inadequate oversight and regulatory loopholes. Given the profound impacts AI can have on society, there's a concern that adaptive approaches might not be thorough enough in assessing and mitigating risks. One fix for this shortcoming is to create layered oversight structures and impact assessment reviews by third-party boards or through extensive use of advisory committee recommendations.\\

Secondly, adaptive methods prioritize speed and agility, which can sometimes come at the cost of in-depth analysis and deliberation. In the context of AI, where decisions can have far-reaching consequences, this could lead to underestimating or overlooking critical issues. One potential fix might be to integrate timed phases of in-depth analysis and public consultation into the agile cycles to ensure comprehensive policy development while preventing discussions which may hinder policy design and implementation.\\

There is also the risk of regulatory uncertainty, the common ambiguity and unpredictability businesses face due to evolving or unclear regulations governing AI development and application. Regulatory uncertainty around innovation can work several ways; for example, for drug development, it typically favors the first mover, but in the case of medical devices, it favors the follow-on entrants \cite{Stern2017}. This uncertainty can stem from frequent changes in policies, inconsistent enforcement, or lack of clarity about compliance requirements and future regulatory directions. A tentative fix is to encourage pilot programs to assess the effectiveness and impact of new policies in controlled environments using regulatory sandboxes \cite{Undheim2022, GonzalezTorres2023}. This measure allows for adjustments based on practical feedback before wider implementation.\\

Another potential solution to this problem is the provision of transparent rationales, timelines and roadmaps for policy changes, allowing businesses (and the public) to plan accordingly. One way is to establish a structured yet flexible framework for policy updates, such as regular review periods. This helps industries anticipate when changes might occur and prepare for them. Additionally, one can implement new regulations in phases, giving industries ample time to adapt; in this context, transition periods where both old and new regulations are temporarily valid could be provided. Governments should further develop clear communication channels and decision-making protocols to streamline consensus-building among diverse stakeholders. However, mass deliberation is often ineffective, demanding, costly, and can result in stalemate or viral ideas that are populist and hard to drive to consensus \cite{Jungherr2023}, dynamics that would need to be managed separately.\\

Finally, regulatory capture is usually thought of as industrial interests being over-proportionately reflected in governance or regulatory initiatives \cite{Tai2017}, often with a profit interest from industry \cite{Saltelli2022}. Generative AI could become an extreme case where not only would big tech try to influence governance, but AI might itself do so, either as part of human-AI hybrids (where human decision-making is augmented by AIs), or if it becomes 'agentic' and has some decision-making and influence capacity of its own. So far, such considerations are highly speculative and may not manifest. But because generative AI development moves fast, we might eventually need to integrate AI systems as independent actors in some form into adaptive governance structures. However, to consider an AI (or a human-AI hybrid) as a distinct actor needing to be governed would challenge the very notion of democracy. Can a machine vote? If it can, how would that work? If it cannot, how can it legitimately be governed? Also, how would we avoid legislative capture given that the AI would to some degree be governing itself? These are all open questions that may provide additional challenges in an adaptive governance framework, but given their speculative nature go beyond our paper. However, we predict that they will require interdisciplinary, cross-actor involvements to find a solution. Setting up flexible, integrative, adaptive governance structures now can help with fostering such discussions among stakeholders, as well as allowing the governance system to adapt should the need arise, no matter how different future iterations of AI turn out to be.

\section{Conclusion}

Because of the speed of development of the technology, its broad scope of application, and its ability to augment human performance and work, generative AI challenges the very notions of governance, trust, and human agency. Additionally, the technology’s capacity to mimic human knowledge work poses a challenge. For these reasons, adaptive governance, where governance and AI co-evolve, maintaining deliberative methods yet increasing speed of adoption, is essential for governing generative AI.\\ 

In this paper, we have presented a comprehensive adaptive AI governance framework that offers a flexible and iterative approach to managing the rapid advancements in artificial intelligence. The framework outlines key actors, shared activities (SCUMIA), and actor-specific activities (FACTI) that are essential for effective AI governance. We have further provided concrete examples of how this framework can be operationalized. Our work aims to support governments in adopting this alternative approach to AI governance; by embracing adaptive governance, governments can create a more agile, inclusive, and responsive regulatory environment that maximizes the potential benefits of AI while mitigating its risks and negative impacts on society.

%It will in all likelihood become difficult and perhaps not fruitful to distinguish between AI-enhanced and manual knowledge work. The legal ramifications alone are perplexing. 

\bibliography{aaai24}

\end{document}